\documentclass[sn-aps,iicol]{sn-jnl}% American PhysicalSociety (APS) Reference Style %referee for peer review / editorial stage
\usepackage{graphicx}%
\usepackage{multirow}%
\usepackage{amsmath,amssymb,amsfonts}%
\usepackage{amsthm}%
\usepackage{mathrsfs}%
\usepackage[title]{appendix}%
\usepackage{xcolor}%
\usepackage{textcomp}%
\usepackage{manyfoot}%
\usepackage{booktabs}%
\usepackage{algorithm}%
\usepackage{algorithmicx}%
\usepackage{algpseudocode}%
\usepackage{listings}%
\usepackage{graphics}
\usepackage[numbers]{natbib}
\usepackage{hyperref}
\usepackage{breqn}
\usepackage{bm}
\usepackage{dcolumn}
\usepackage[modulo]{lineno}
\newcommand{\revisedJP}[1]{{ #1}} % remove red color at once for all revisions

\graphicspath{figs/}

\begin{document}

\title[Deuteron TMDs]{Experimental Study of Tensor Structure Function of Deuteron}


%\author*[1]{\fnm{fname} \sur{surname}}\email{abc@jlab.org}

\author*[1]{\fnm{Jiwan} \sur{Poudel}}\email{jpoudel@jlab.org}
\author[2,3]{\fnm{Alessandro} \sur{Bacchetta}}
\author[1]{\fnm{Jian-Ping} \sur{Chen}}%\email{jpchen@jlab.org}
\author[4]{\fnm{Nathaly} \sur{Santiesteban}}

%\affil[]{\orgname{Jefferson Lab}, \orgaddress{\street{12000 Jefferson Ave}, \city{Newport News}, \postcode{23606}, \state{VA}, \country{United States}}}

\affil*[1]{\orgname{Jefferson Lab}, \orgaddress{\street{12000 Jefferson Ave}, \city{Newport News}, \postcode{23606}, \state{VA}, \country{United States}}}

%\affil[2]{\orgname{University of Pavia}, \orgaddress{\street{Via Bassi 6}, \postcode{I-27100}, \state{Pavia}, \country{Italy}}}

%\affil[2]{\orgname{Dipartimento di Fisica "Alessandro Volta", Università di Pavia}, \orgaddress{\street{Via Bassi}, \postcode{6 - 27100}, \state{Pavia}, \country{Italy}}}
\affil[2]{\orgname{Università di Pavia}, \orgaddress{\street{Via Bassi}, \postcode{6 - 27100}, \state{Pavia}, \country{Italy}}}

\affil[3]{\orgname{INFN Pavia}, \orgaddress{\street{Via Agostino Bassi}, \postcode{6 - 27100}, \state{Pavia}, \country{Italy}}}

\affil[4]{\orgname{University of New Hampshire}, \orgaddress{\street{9 Library Way}, \city{Durham}, \postcode{03824}, \state{NH}, \country{United States}}}

%\affil[3]{\orgdiv{}, \orgname{}, \orgaddress{\street{}, \city{}, \postcode{}, \state{}, \country{}}}


\abstract{The deuteron is the lightest spin-1 nucleus, consisting of a weakly bound system of two spin-$\frac{1}{2}$ nucleons. One intriguing characteristic of the deuteron is the tensor polarized structure, which cannot be naively constructed combining the proton and neutron structure. The tensor structure of the deuteron provides unique insights into the quarks and gluons distributions and their dynamics within the nucleus. It can be studied experimentally through inclusive and semi-inclusive Deep Inelastic Scattering (DIS) of electrons on tensor polarized deuterons. One-dimensional (longitudinal-momentum-dependent) tensor structure functions are extracted from the inclusive DIS, whereas three-dimensional with additional transverse-momentum-dependent tensor structure functions are extracted from the semi-inclusive DIS. Experimentally, achieving high tensor polarization for such measurements has been a challenge. Significant progress has recently been made in enhancing the tensor polarization for polarized deuteron target, opening up a new window for experimental studies of the deuteron tensor structure. In this article, we discuss the tensor structure functions of the deuteron and the experimental schemes to extract these functions at Jefferson Lab, highlighting the potential measurements of the transverse-momentum-dependent tensor structure functions.}

\keywords{ deuteron, tensor structure, tensor polarization}

\maketitle

\section{Introduction}\label{sec:1}

The deuteron is the simplest spin-1 nucleus, which consists of a weakly bound system of spin-$\frac{1}{2}$ proton and neutron. Historically, the deuteron has been used to extract neutron distribution functions. However, there is growing interest in expanding its use to understand the full spin structure of the bound system. The tensor nature of the spin-1 system is missing completely in spin-$\frac{1}{2}$ particles, while the spin-1 deuteron has often been approximated as a simple combination of a spin-$\frac{1}{2}$ proton and a neutron. Pioneering work by Hoodbhoy, Jaffe, and Manohar described the tensor structure functions of a spin-1 target~\cite{Hoodbhoy:1988am}, with a particular focus on the leading twist contributions that could be measured with a tensor-polarized target. Theoretical interest in the structure functions that describe the spin-1 systems has increased over the years, with  several intriguing mysteries surrounding the spin-1 structure of the deuteron~\cite{Nikolaev:1996jy,Bora:1997pi,Umnikov:1996qv,Edelmann:1997qe}. Further investigation on the tensor structure function can open a unique window to study spin physics in light nuclei. The tensor structure functions enable the study of the interaction between partons beyond what can be determined with spin-$\frac{1}{2}$ nucleon structure functions alone~\cite{PhysRevD.82.2010_Kumano,proceeding_Kumano_2022,Miller_2014}. Measurement of these tensor structure functions enlightens the exotic states in light nuclei, which will ultimately provide new information on Quantum Chromo-Dynamics (QCD) and the partonic structure of nuclei.

%While the spin-$\frac{1}{2}$ nature of the nucleons has been studied extensively in experiments, the tensor component of spin-1 systems has rarely been explored experimentally to unravel the mysteries. %%%SNS:commented because this phrase is already two times in the text

The only DIS tensor measurement to date was the inclusive tensor structure function $b_1$ by the HERMES Collaboration, published in 2005~\cite{PhysRevLett.95.2005_Hermes}. The HERMES data revealed a significantly larger value of $b_1$ at moderate $x\geqslant 0.2$, albeit with relatively large uncertainty. All conventional nuclear physics models, including effects beyond a free proton and neutron, predict $b_1$ to be very small ($<10^{-3}$) in this region. Therefore, a comprehensive understanding of the tensor structure of the spin-1 deuteron is required for understanding the partonic structure of light nuclei. While the one-dimensional $x$-dependent tensor distribution functions such as $b_1$ can be accessed via inclusive processes, we can also access tensor Transverse-Momentum-Dependent (TMD) distribution functions via Semi-Inclusive DIS (SIDIS) processes which offer novel tools for investigating hadron tomography in momentum space as described theoretically in Ref.~\cite{PhysRevD.62.2000_bacchetta,PhysRevD.103.2021_Kumano,Cosyn:2020kwu}. \revisedJP{Various model calculations with convolution and covariant approaches are available to explain the one-dimensional $b_1(x)$ tensor structure function of the deuteron as in Ref.~\cite{Hoodbhoy:1988am,PhysRevD.82.2010_Kumano,Miller_2014,Cosyn:2017fbo,Umnikov:1996qv}, but there is only one published model calculation with the tensor TMD formulation on Spin-1 system. It is a framework with the covariant formulation for the leading twist-TMDs of the Spin-1 system which has been used in the model calculation of the $\rho^+$ meson as described in Ref.~\cite{PhysRevC_Cloet_2017}. A qualitative comparison of this model with the existing $b_1(x)$ data on deuteron is discussed in this article, but detailed calculations of TMDs on deuteron using this model is being worked on and coming out in few months~\cite{Cloet_private_2025}. Even the result of spin-1 TMDs of $\rho^+$ meson are interesting for the tensor polarized components, as they don't appear in spin-$\frac{1}{2}$ case and have interesting features about $x=\frac{1}{2}$~\cite{PhysRevC_Cloet_2017}. Validation and verification of all these theoretical models for $b_1$ as well as tensor polarized TMDs require lots of experimental data, and Jefferson Lab is one of the appropriate facility to study these tensor polarized distribution functions experimentally.}

The spin-1 system such as the deuteron has both vector polarization ($\mathcal{P} \equiv n_{+} - n_{-}$) and tensor polarization ($\mathcal{Q} \equiv n_{+} + n_{-}-2n_0$), depending on the population densities ($n_{+}, n_{0}, n_{-}$) of the three different magnetic sub-states ($m = +1, 0, -1$) when placed in a magnetic field. Dynamic Nuclear Polarization (DNP) method has been utilized for a long time to polarize the solid nuclear targets which could produce significantly large vector polarization.
Using DNP in solid targets such as ND$_3$, the vector polarization of the deuteron can reach up to 50\%, which corresponds to 19\% tensor polarization in thermal equilibrium according to Boltzmann statistics. With reasonable electron beam current, average in-beam vector polarization can reach 30-40\%, which corresponds to 7-12\% tensor polarization.  This level of tensor polarization is not high enough to perform precision experiments, and the polarized target community has been striving to achieve higher polarization levels in recent years~\cite{Keller:2020wan,Clement:2023eun} ensuring the future tensor polarized experiments with higher luminosity. Current efforts show the tensor polarization can be enhanced to over 30\%~\cite{Keller:2020wan}. The recent progress in the development of high tensor polarization opens up a window of opportunities for further investigation into the spin-1 tensor structure functions with dedicated tensor experiments.

\revisedJP{The tensor experimental program at Jefferson lab counts with an approved $b_1$ experiment in Hall-C~\cite{proposal_b1_jlab} to measure the tensor structure function with better precision around the cross-over region of HERMES data ($0.2 < x < 0.5$).} Additionally, recent interest to measure the tensor contributions in the SIDIS processes has motivated us for the tensor proposals with potentials of getting information on the transverse-momentum-dependent tensor structure functions~\cite{CAA_tensor_arxiv,LOI_pac52_spin1}. First, there will be an exploratory study of the tensor contribution using Cebaf Large Acceptance Spectrometer (CLAS12) data, which was optimized for vector measurements but still retains average tensor polarization of around 10\%. Then, proposal for dedicated experiments with enhanced tensor polarized target and experimental setup focused on SIDIS measurements will be submitted to Jefferson Lab. Spectrometers in Hall-C could provide higher precision result at particular kinematics, whereas large acceptance devices such as Solenoinal Large Intensity Device (SoLID) in Hall-A~\cite{SoLID_whitePaper2024} and CLAS12 in Hall-B~\cite{CLAS12_BURKERT} could provide information on broad kinematic region suitable for multidimensional studies. This article discusses the formalism of tensor structure functions for both inclusive and SIDIS processes in Sec.~\ref{sec:2}. Section~\ref{sec:3} describes the experimental observables for accessing the tensor contributions, while Sec.~\ref{sec:4} details the experiments that enable the extraction of tensor structure functions at Jefferson Lab, with a focus on the novel and yet-to-be-measured SIDIS processes. Section~\ref{sec:5} concludes this article with a summary.

\section{Formalism of Tensor Structure Function}\label{sec:2}

The differential cross-section for the lepton-hadron inelastic scattering mechanism can be expressed as
\begin{align}
%    \sigma &\propto L_{\mu\nu}W^{\mu\nu}\\
%    \frac{d^2\sigma}{d\Omega dE'} &= \frac{\alpha^2}{q^4}\frac{E'}{E}L_{\mu\nu}W^{\mu\nu}\\
    \frac{d^2\sigma}{dx dy} &= \frac{2\pi y\alpha^2M}{Q^4} L_{\mu\nu}W^{\mu\nu}
%\frac{d^2\sigma}{dx dy} = x (s-M^2) \frac{d^2\sigma}{dx dQ^2} = %\frac{2\pi M \nu}{E'} \frac{d^2\sigma}{d\Omega dE'} 
    %\frac{d^2\sigma}{dx~dQ^2} &= \frac{2\pi y\alpha^2}{xQ^4(s-M^2)} L^{\mu\nu}W_{\mu\nu}
    \label{eq:cross-section}
\end{align}
where $L_{\mu\nu}$ and $W^{\mu\nu}$ are the leptonic and hadronic tensor respectively. We consider the four momentum of initial state lepton as $l\equiv(E_l,\Vec{p_l})$, the final state lepton as $l'\equiv(E'_l,\Vec{p'_l})$, the initial state hadron as $P\equiv(E_N,\Vec{p}_N)$ and the exchanged photon $q\equiv (E_l - E'_l, \Vec{p_l} - \Vec{p'_l})$. Now, the kinematic variables can be expressed as, $Q^2=-q^2 >0$, $x=\frac{Q^2}{2P\cdot q}$, $y=\frac{P\cdot q}{P\cdot l}$ and $s=(l+P)^2$. Here, $M$ represents the rest mass of hadron and $\alpha$ is fine structure constant.

The following subsections describe the tensor structure functions in detail for both inclusive and semi-inclusive DIS, in which the SIDIS process will be discussed only for the case where the polarization is along the electron beam, i.e., longitudinally polarized.

\subsection{Inclusive Process}\label{sec:2a_inclusive}
In the inelastic scattering of a lepton from a Spin-1 target, the hadronic-tensor in Eq.~(\ref{eq:cross-section}) can be expressed in terms of structure functions as described in Ref.~\cite{Hoodbhoy:1988am, PhysRevD.42.1990_CloseKumano}
\begin{align}{\label{eq:inclusive_hadronTensor}}
    W_{\mu\nu}=&-F_1g_{\mu\nu}+F_2\frac{P_\mu P_\nu}{\nu}+\frac{ig_1}{\nu}\epsilon_{\mu\nu\lambda\sigma}q^\lambda s^\sigma  \nonumber
    \\&+\frac{ig_2}{\nu^2}\epsilon_{\mu\nu\lambda\sigma}q^\lambda (P\cdot q s^\sigma-s\cdot q P^\sigma) \nonumber
    \\&-b_1r_{\mu\nu}+\frac{b_2}{6}(s_{\mu\nu}+t_{\mu\nu}+u_{\mu\nu}) \nonumber
    \\&+\frac{b_3}{2}(s_{\mu\nu}-u_{\mu\nu})+\frac{b_4}{2}(s_{\mu\nu}-t_{\mu\nu})
\end{align}
where $F_{1,2}$, $g_{1,2}$ and $b_{1,2,3,4}$ are the unpolarized, polarized and inclusive tensor structure functions respectively. Here, $r_{\mu\nu}$, $s_{\mu\nu}$, $t_{\mu\nu}$ and $u_{\mu\nu}$ are some variables used in Ref.~\cite{Hoodbhoy:1988am}, which depend on the reaction kinematics. %we might not need following lines in this article, only naive model could be enough
%With the QCD factorization scale $\mu$ and the renormalized quark distribution $q(x,\mu^2)$, the unpolarized structure function can be expressed under the improved parton model as
%\begin{align}
%    F_1 =& \sum_{q,\Bar{q}}e_q^2 \int^1_x\frac{d\xi}{\xi} q(\xi,\mu^2) \bigg[ \delta(1-\frac{x}{\xi})+\frac{\alpha_s}{2\pi}S(x/\xi)\ln{\frac{Q^2}{l^2}}+...\bigg]
%\end{align}
%where $\xi$ momentum fraction carried by quark and $S(x/\xi)$ is the splitting function given by $S(z)=\frac{4}{3}\frac{1+z^2}{1-z}$.

In the naive parton model, the structure functions can simply be expressed in leading order terms as:
\begin{align}
    F_1 =& \frac{1}{2}\sum_{q,\Bar{q}} e_q^2 [q (x)] \nonumber\\
    g_1 =& \frac{1}{2}\sum_{q,\Bar{q}} e_q^2[\Delta q(x)] \nonumber\\
    b_1 =& \frac{1}{2}\sum_{q,\Bar{q}} e_q^2[\delta q (x)]
\end{align}
where $q(x)=\frac{2}{3}[q_{\uparrow}^0 (x)+q_{\uparrow}^1 (x)+q_{\uparrow}^{-1} (x)]$, $\Delta q(x) = [q_{\uparrow}^1 (x)-q_{\downarrow}^1 (x)]$ and $\delta q(x) = 2q_{\uparrow}^0 (x) - [q_{\uparrow}^1(x)+q_{\downarrow}^1 (x)]$, with the summation over all the flavour of quarks $q$ and antiquarks $\Bar{q}$. Here, $q_{\uparrow (\downarrow)}^m$ represents the probability of finding a quark with momentum fraction $x$ and spin up (down) in the spin-1 hadron with helicity $m$. Due to  parity invariance, it follows $q_\uparrow^m = q_\downarrow^{-m}$.

Tensor structure functions $b_1$ and $b_2$ are leading twist terms, whereas $b_3$ and $b_4$ are higher twist terms. The leading twist terms also satisfies the Callan-Gross type relation, $b_2=2xb_1$. We can also express the spin-averaged distributions as $q^0 (x)=q_{\uparrow}^0 (x) + q_{\downarrow}^0 (x) = 2q_{\uparrow}^0 (x)$ and $q^1 (x)=q_{\uparrow}^1 (x) + q_{\downarrow}^1 (x)$. As a result, $\delta q(x) = q^0 (x) - q^1(x)$, which implies that $b_1$ depends on the spin average quark distributions inside deuteron. Therefore, the tensor structure function $b_1$ can be interpreted as the difference in partonic distribution in different helicity states ($m=0~\text{and}~1$) of the spin-1 hadron.

\subsection{SIDIS Process}\label{sec:2b_sidis}

The total cross section in the single photon exchange process of the electron-deuteron (spin-1) interaction, $e(l)+d(P)\rightarrow e(l')+h(P_h)+X$, where \revisedJP{at least} an additional hadron is detected in the final state, for a longitudinally polarized target can be expressed in terms of transverse momentum-dependent structure functions as in Ref.~\cite{CAA_tensor_arxiv,LOI_pac52_spin1}, 
\begin{align}
    &\frac{d\sigma}{dx~dy~d\psi~dz~d\phi_h~dP^2_{h\perp}} \nonumber
    \\ & = \frac{y\alpha^2}{2(1-\epsilon)xQ^2}\left(1+\frac{\gamma^2}{2x}\right) \Bigg[F_{UU,T}+\epsilon F_{UU,L} \nonumber
    \\ &+\sqrt{2\epsilon(1+\epsilon)}\cos{\phi_h} F_{UU}^{\cos\phi_h}+\epsilon \cos(2\phi_h)F_{UU}^{\cos{(2\phi_h)}} \nonumber
    \\ &+\lambda_e\sqrt{2\epsilon(1-\epsilon)}\sin{\phi_h}~F_{LU}^{\sin{\phi_h}}\nonumber\\
%\end{align}
%    \hline
%    &-------------------------\nonumber\\
%\begin{align}
    &+\revisedJP{S_\parallel}\bigg\{\sqrt{2\epsilon(1+\epsilon)}\sin{\phi_h} F_{UL}^{\sin{\phi_h}}+\epsilon\sin{(2\phi_h)} F_{UL}^{\sin{2\phi_h}}\nonumber\\
    &+\lambda_e\bigg(\sqrt{1-\epsilon^2}~F_{LL}+\sqrt{2\epsilon(1-\epsilon}\cos{\phi_h}~F_{LL}^{\cos{\phi_h}}\bigg)\bigg\}\nonumber\\
%    &-------------------------\nonumber\\
    &+\revisedJP{T_{\parallel\parallel}}\bigg\{F_{U(LL),T}+\epsilon F_{U(LL),L} \nonumber
    \\&+\sqrt{2\epsilon(1+\epsilon)}\cos{\phi_h}~F_{U(LL)}^{\cos{\phi_h}} +\epsilon\cos{(2\phi_h)}~F_{U(LL)}^{\cos{2\phi_h}} \nonumber
    \\&+\lambda_e\sqrt{2\epsilon(1-\epsilon)}\sin{\phi_h}~F_{L(LL)}^{\sin{\phi_h}}
    \bigg\}
    \Bigg]
    \label{eq:cross-section}
\end{align}

\noindent where \revisedJP{$\lambda_e=\pm1$, depending on the} electron helicity, $\epsilon$ is a \revisedJP{parameter of the virtual-photon polarization (defined as a ratio of longitudinal and transverse photon flux)}. Here, $F$ represents the structure functions where first letter in subscript denotes the beam polarization, second (and third also if enclosed inside the parentheses) indicates the target polarization and the last one after the separator comma denotes the virtual photon polarization.

\revisedJP{In addition, $S_\parallel$ is the vector polarization, and $T_{\parallel\parallel}$ is the tensor polarization of target in parallel to the virtual photon direction which are related with $\mathcal{P}$ and $\mathcal{Q}$ along the direction of electron beam as
$S_{\parallel} = \cos{\theta_{ql}~\mathcal{P}}$ and 
$T_{\parallel\parallel} = \frac{1+3\cos(2\theta_{ql})}{4\sqrt{6}}\mathcal{Q}$ where $\theta_{ql} = \arcsin\big(\gamma\sqrt{\frac{1-y-y^2\gamma^2/4}{1+\gamma^2}}\big)$ is the angle between virtual photon and electron beam direction as described in Ref.~\cite{Diehl_2005_CoodinateFrame,Cosyn:2017fbo}. Other transverse components ($T_{\parallel\perp}$ and $T_{\perp\perp}$) do not vanish while considering target polarization along the beam direction, but they are suppressed by at least a factor $\gamma=\frac{2xM}{Q}$, which is small in the DIS region. In the Bjorken scaling limit, $\gamma\rightarrow0$ and $\theta_{ql}\rightarrow0$, and the cross-section expressed has comparatively small transverse contribution which is omitted in Eq.~\ref{eq:cross-section}.}

The total cross section of a vector and tensor-polarized deuteron target is composed of unpolarized, vector and tensor contributions. In Eq.~\ref{eq:cross-section}, the first three lines represent the unpolarized contribution. Next two lines within the curly brackets multiplied by $S_{\parallel}$ represent the contribution from vector polarization, weighted by the amount of vector polarization $S_{\parallel}$. Lastly, the tensor contribution is weighted by the size of tensor polarization, $T_{\parallel\parallel}$, appears in the last three lines.

The tensor polarized structure functions that are extracted experimentally can be further expressed in terms of the TMD distribution functions ($f,h,e~\text{and}~g$) as a convolution with the fragmentation correlation functions ($D,H,E~\text{and}~G$) following the TMD factorization schemes as~\cite{CAA_tensor_arxiv,LOI_pac52_spin1}:
\begin{subequations} \label{eq:TMD_observable}
    \begin{align}
    {F_{U(LL),T}} =& C[f_{1LL}D_1] \label{eq:TMD_observable1}\\
    {F_{U(LL),L}} =& 0 \label{eq:TMD_observable2}\\
    {F_{U(LL)}^{\cos{\phi_h}}} =& \frac{2M}{Q}C\Bigg[-\frac{\hat{\mathbf{h}}\cdot \mathbf{k_T}}{M_h}\bigg(xh_{LL}H_1^\perp \nonumber
    \\&+\frac{M_h}{M}f_{1LL}\frac{\Tilde{D}^\perp}{z}\bigg) -\frac{\hat{\mathbf{h}}\cdot \mathbf{p_T}}{M}\bigg(xf_{LL}^\perp D_1 \nonumber
    \\&+\frac{M_h}{M}h_{1LL}^\perp\frac{\Tilde{H}}{z}\bigg)\bigg] \label{eq:TMD_observable3}\\
    {F_{U(LL)}^{\cos{2\phi_h}}} =& C\Bigg[-\frac{2(\hat{\mathbf{h}}\cdot \mathbf{k_T})(\hat{\mathbf{h}}\cdot \mathbf{p_T})-\mathbf{k_T}\cdot \mathbf{p_T}}{MM_h} \nonumber
    \\ &\times h_{1LL}^{\perp} H_1^{\perp} \bigg] 
    \label{eq:TMD_observable4} \\
    {F_{L(LL)}^{\sin{\phi_h}}} =& \frac{2M}{Q}C\Bigg[-\frac{\hat{\mathbf{h}}\cdot \mathbf{k_T}}{M_h}\bigg(xe_{LL}H_1^\perp \nonumber
    \\&+\frac{M_h}{M}f_{1LL}\frac{\Tilde{G}^\perp}{z}\bigg) +\frac{\hat{\mathbf{h}}\cdot \mathbf{p_T}}{M}\bigg(xg_{LL}^\perp D_1 \nonumber
    \\&+\frac{M_h}{M}h_{1LL}^\perp\frac{\Tilde{E}}{z}\bigg)\bigg] \label{eq:TMD_observable5}
    \end{align}
\end{subequations}
where $p$ is the momentum of quark coming out of the target, $k$ is momentum of the quark decaying into the outgoing hadron after the interaction with virtual photon $q$, and $\hat{\mathbf{h}}=\frac{\mathbf{P}_{h\perp}}{|{\mathbf{P}_{h\perp}}|}$ as described in Ref.~\cite{PhysRevD.62.2000_bacchetta,A_Bacchetta_2007}.

In TMD distribution functions e.g. $f_{1LL}$, the numeric subscript at the beginning is assigned for twist (1 for twist-2) whereas alphabetic subscript are assigned for the hadron polarization. However, twist-3 TMDs have no number subscript following the notations from Ref.~\cite{A_Bacchetta_2007} and~\cite{PhysRevD.103.2021_Kumano}. Integration of TMD distribution functions over the partonic transverse momentum $p_T$ provide the collinear distribution functions. Among the various collinear functions obtained from TMDs, $f_{1LL}$ is related with the inclusive distribution function $b_1$.
%and with an arbitrary funtion $w(\mathbf{p}_T,\mathbf{k}_T$), we have
%\begin{align}
%    C =&~x\sum_{a}^{} e_a^2 \int d^2\mathbf{p}_T~d^2\mathbf{k}_T~\delta^2(\mathbf{p}_T - \mathbf{k}_T - \mathbf{P}_{h\perp}/z)~w(\mathbf{p}_T,\mathbf{k}_T)~f^a(x,p_T^2)~D^a_h(z,k_T^2)
%\end{align}
%with sum over quarks and anti-quarks~\cite{A_Bacchetta_2007}. 

\section{Tensor Observables}\label{sec:3}

Considering the deuteron target with vector polarization ($\mathcal{P}$) and tensor polarization ($\mathcal{Q}$), the cross-section ($\sigma$) of deuteron can be written in terms of different asymmetries ($A$) as described in Ref.~\cite{Arenhovel:1988qh,Leidemann1991PRC}:
\begin{align}\label{eq:cross-section-assymetry}
    \sigma =& \sigma_U \bigg[1+ \mathcal{P} A^V + \mathcal{Q} A^T \nonumber
    \\&+\lambda_e(A^e + \mathcal{P} A^{eV} + \mathcal{Q} A^{eT})\bigg]
\end{align}
where $\sigma_U$ is the unpolarized cross-section, $\lambda_e$ is the beam helicity, and the electron helicity assymetry $A^e = \frac{1}{2\lambda_e\sigma_U}\big[\sigma(\lambda_e, \mathcal{P} =0, \mathcal{Q}=0)-\sigma(-\lambda_e, \mathcal{P} =0, \mathcal{Q}=0)\big]$ which provide information about the fifth structure function with an out-of-plane measurement. Similarly, $A^V$ and $A^{T}$ are vector and tensor target asymmetry respectively, which could be accessed by:
\begin{align}
    A^V =& \frac{1}{2\mathcal{P}\sigma_U}\bigg[\sigma(\lambda_e=0,\mathcal{P},\mathcal{Q})-\sigma(\lambda_e=0,-\mathcal{P},\mathcal{Q})\bigg] \\
    A^T =& \frac{1}{2\mathcal{Q}\sigma_U}\bigg[\sigma(\lambda_e=0,\mathcal{P},\mathcal{Q})+\sigma(\lambda_e=0,-\mathcal{P},\mathcal{Q}) \nonumber
    \\&- 2\sigma_U\bigg]
    \label{eq:assymmetry2}
\end{align}
Lastly, $A^{eV}$ and $A^{eT}$ are the vector and tensor beam-target double asymmetry respectively.

Accessing the tensor asymmetry requires spin manipulations to suppress the contributions from other asymmetries. Ideally, using an unpolarized electron beam will cause the electron-related asymmetries to vanish. Alternatively, a weighted average of the beam polarization can also suppress these contributions. The extraction of the tensor asymmetry, along with the related physics observables, is described below for both inclusive and SIDIS processes.

\subsection{Tensor Asymmetry in Inclusive Processes}\label{subsec:3a_inclusive}

Considering an unpolarized electron beam, the inclusive cross-section can be expressed as
\begin{align}\label{eq:inclusive_sigma}
    \sigma =&\sigma_U \bigg(1 + \mathcal{Q} A^{T}\bigg)
\end{align}
where electron related asymmetries in Eq.~\ref{eq:cross-section-assymetry} vanish because of an unpolarized beam ($\lambda_e=0$). \revisedJP{Equation~\ref{eq:inclusive_sigma} is valid for polarized beam if we average the equally distributed positive and negative electron helicity events.} The vector asymmetry ($A^V$) also vanishes for the inclusive process with single photon exchange approximation \revisedJP{due to parity invariance}.

The inclusive tensor asymmetry can then be written from the above expression as,
\begin{align}\label{eq:inclusive_AT}
    A^T =& \frac{1}{\mathcal{Q}}\bigg(\frac{\sigma}{\sigma_U}-1\bigg)
\end{align}
%where $\sigma = \sigma(\lambda_e=0,\mathcal{P},\mathcal{Q}) +   \sigma(\lambda_e=0,\mathcal{-P},\mathcal{Q})$, and to reduce systematical drifts, $\sigma_U = \sigma(\lambda_e=0,\mathcal{P},0) +   \sigma(\lambda_e=0,\mathcal{-P},0)$, which reduces the spin-up and spin-down times.

The experiments are not performed with a pure deuterium target; instead, the target material $ND_3$ is held in cups and submerged in a bath of $^4$He. This \revisedJP{polarized target }setup causes the incident electron beam to interact with different materials \revisedJP{in addition to the deuteron at the target}, which can influence the intended measurement. Accounting for this contamination is achieved using the dilution factor ($f$), which is  a fraction of counts that comes from the deuteron ($D$) to the overall counts from all the material in the target: $f={n_D\sigma_D}/{\sum_i n_i \sigma_i}$. Here, $n_i$ is the number of nuclei of particular material in beam  (e.g., deuteron, nitrogen, helium bath etc.), each with scattering cross-section $\sigma_i$. After accounting for this correction and assuming that acceptance is the same in the cross-section ratios, the asymmetry can be expressed as
\begin{align}\label{eq:AT_measured}
    A^T =& \frac{1}{f \mathcal{Q}}\bigg(\frac{N_T}{N_U}-1\bigg)
\end{align}
where $N_T$ is the tensor polarized yield and $N_U$ is the unpolarized yield. Both of these yields should be charge normalized and efficiency corrected before getting the asymmetry. 

Furthermore, the inclusive tensor asymmetry can also be extracted using Eq.~\ref{eq:assymmetry2} which requires tensor polarization yield on both positive and negative vector polarity. \revisedJP{Availability of data on both target polarity allows to estimate the false asymmetry by analyzing the yield on sum and difference between two different target polarity, which ultimately helps to correct the experimental tensor asymmetry. In addition, asymmetry obtained from Eq.~\ref{eq:inclusive_sigma} and~\ref{eq:inclusive_AT} are based on the single photon approximation, but the result of Eq.~\ref{eq:assymmetry2} is without any approximations which would be suitable to investigate the contribution from higher order terms.}

The tensor asymmetry $A^T$ obtained from the inclusive experiments contains information about all the tensor structure functions  ($b_1,b_2,b_3,b_4$) presented in Eq.~\ref{eq:inclusive_hadronTensor}. Reference~\cite{Cosyn:2017fbo} details the relation of tensor asymmetry with the tensor structure functions, including the contribution from higher twist terms. In the case of deuteron polarization parallel \revisedJP{ to the electron beam direction (not the virtual photon direction), additional transverse components of tensor polarization parameters also appears in the cross-section due to $\theta_{ql}$ as described in Sec.~\ref{sec:2}. While considering the Bjorken scaling limit with fixed $x$, large $Q^2$ and small $\gamma$, the transverse polarization contribution is very small and can be omitted. Additionally,  tensor structure functions follow the Callan-Gross type relation ($b_2 = 2xb_1$) and the higher twist terms can be neglected in the Bjorken limit. In this kinematic limit,} the tensor asymmetry can be expressed in terms of leading order terms only, as below:
\begin{align}{\label{eq:hermes_Azz}}
    A^T \approx -\frac{1}{3} \frac{b_1}{F_1^d}
\end{align}
where $F_1^d$ is the leading twist structure function of deuteron. \revisedJP{The inclusive tensor assymetry $A^T$ used in this paper is equivalent to $A_{zz}/2$ defined in HERMES Collaboration~\cite{PhysRevLett.95.2005_Hermes} and JLab $b_1$ Collaboration~\cite{proposal_b1_jlab}. HERMES collaboration used this leading order expression in Eq.~\ref{eq:hermes_Azz} to calculate the $b_1$ structure function.} Using the available world data of $F_1^d$, tensor structure function $b_1$ can be extracted from the experimentally measured tensor asymmetry $A^T$ using above expression. However, later studies show that neglecting higher twist effects may not be trivial and should not be ignored at low $Q^2$~\cite{Cosyn:2017fbo}. \revisedJP{The 11 GeV beam energy at Jefferson Lab provides data on modest $Q^2$ region, and the upgraded beam energy of 22 GeV at Jefferson Lab and the EIC can provide data on higher $Q^2$, minimizing the uncertainty contribution from the neglected terms.}

\subsection{Tensor Asymmetry in SIDIS Processes}

Considering an unpolarized electron beam, the SIDIS cross-section can be written as
\begin{align}\label{eq:sidis-assymetry}
    \sigma =& \sigma_U \bigg(1+ \mathcal{P} A^V + \mathcal{Q} A^{T}\bigg)
\end{align}
where the terms containing beam polarization ($\lambda_e$) in Eq.~\ref{eq:cross-section-assymetry} vanish for an unpolarized beam.

As in inclusive process, considering $f_D$ as the dilution factor, the experimental tensor asymmetry of deuteron is calculated from Eq.~\ref{eq:assymmetry2} as
\begin{align}
   A^T =& \frac{1}{2f_D \mathcal{Q}}\bigg(\frac{N_T^{+\mathcal{P}}}{N_U}+\frac{N_T^{-\mathcal{P}}}{N_U}-2\bigg)
\end{align}
where $N_T^{+\mathcal{P}}$ is the yield for the tensor polarized target with positive vector polarization, whereas $N_T^{-\mathcal{P}}$ is the yield with negative vector polarization. All of these yield should be charge normalized and efficiency corrected while getting the asymmetry.

The tensor asymmetry obtained experimentally via the SIDIS process is related with Eq.~\ref{eq:cross-section}, and it contains the information of the transverse-momentum-dependent tensor structure functions in Eq.~\ref{eq:TMD_observable}. The tensor asymmetry can be expressed deducing from Eq.~\ref{eq:cross-section} and~\ref{eq:assymmetry2} as
\begin{align}\label{eq:tensor_assymtry_Rel}
    A^T \sim & \revisedJP{\frac{1}{\sigma_U}}\bigg[F_{U(LL),T}+\sqrt{2\epsilon(1+\epsilon)}\cos{\phi_h}~F_{U(LL)}^{\cos{\phi_h}} \nonumber
    \\&+\epsilon\cos{(2\phi_h)}~F_{U(LL)}^{\cos{2\phi_h}}\bigg]
\end{align}
And to extract the different tensor structure functions $F_{U(LL)T}$, $F_{U(LL)}^{\cos{2\phi_h}}$ and $F_{U(LL)}^{\cos{\phi_h}}$, azimuthal angular modulations can be used. \revisedJP{These measured structure functions can then be used in Eq.~\ref{eq:TMD_observable} to calculated the tensor TMDs. While considering the target polarization $\mathcal{Q}$ along the electron beam direction experimentally, Eq.~\ref{eq:tensor_assymtry_Rel} is valid approximation for the kinematic region as described in Sec.~\ref{sec:2b_sidis}.}

\section{Experimental Programs on Tensor Structure Functions}\label{sec:4}

Interest in the experimental studies of tensor structure functions at Jefferson Lab has been rising. One experiment has already been approved to measure the inclusive $b_1$ structure function, and there is a CLAS12 Approved Analysis (CAA) proposal, along with a Letter of Intent for a dedicated experiment in Hall-C, to study the tensor TMD structure function through SIDIS processes. Expansion of these measurements in the future for the multidimensional studies of tensor TMDs using large acceptance spectrometer such as SoLID is under discussion. The following sub-sections provide a description of these programs, with particular focus on the pioneering SIDIS measurement.

\subsection{Inclusive Programs}

\begin{figure*}[htb]
    \centering
    \includegraphics[height=8cm,width=12cm]{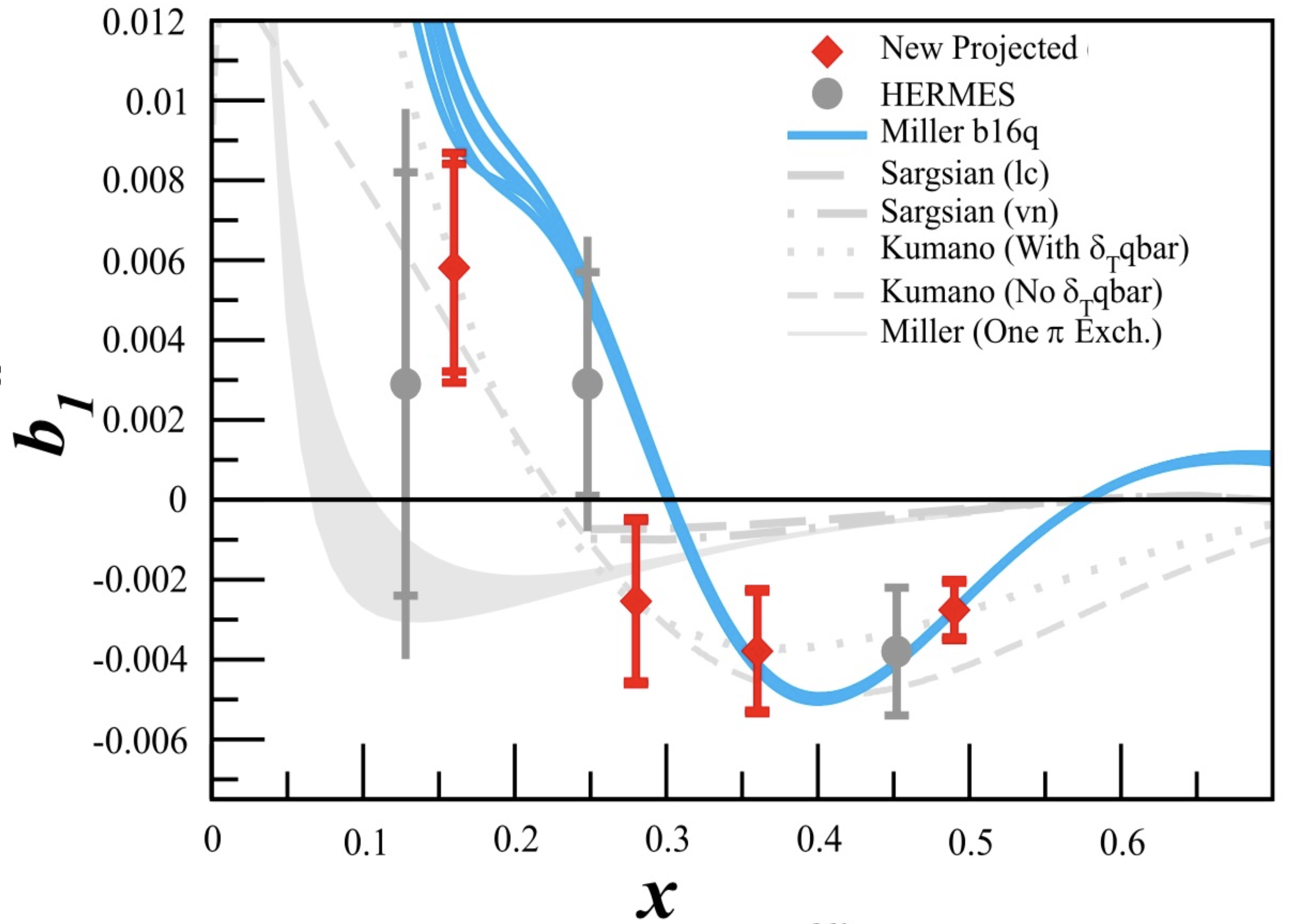}
    \caption{Projected result of $b_1$ experiment proposed at Hall-C of Jefferson Lab (red) in comparing with the available HERMES data and various theoretical model predictions within the Bjorken-x limit $0.1<x<0.7$. \revisedJP{Error bars are reestimated during the PAC51 Jeopardy presentation considering 85 nA electron beam incident on $26\%$ enhanced tensor polarized target for 36 PAC days}~\cite{proposal_b1_jlab_jeopardyPAC51}.}
    \label{fig:b1_projected}
\end{figure*}

The inclusive tensor structure function $b_1$ of the deuteron was first measured experimentally by the HERMES Collaboration~\cite{PhysRevLett.95.2005_Hermes} in 2005, with a positron beam impinging on a tensor polarized deuteron gas target. %The non-zero value of $b_1$ as shown in Fig.~\ref{fig:b1_projected} and the violation of Close-Kumano sum rule in the HERMES result demanded new approach to fit those data, for example: the tensor polarized anti-quark distribution as explained in Ref.~\cite{PhysRevD.82.2010_Kumano} or hidden-color six quark model as explained in Ref.~\cite{Miller_2014}. 
There is an ambiguity to interpret the HERMES result because of the larger uncertainty, as all the data points are comparable to zero at 2-sigma level. Further experimental studies with improved precision are demanded, and a new experiment is approved to measure the inclusive tensor structure function $b_1$ in the Hall-C of Jefferson lab~\cite{proposal_b1_jlab}. This experiment utilizes the existing HMS and SHMS spectrometers which measures $b_1$ with higher precision \revisedJP{in the kinematic region of $0.16 < x < 0.49$, $0.8 < Q^2 < 5.0$ GeV$^2$ and $W \geq 1.85$ GeV, and map out the $x$-region where a zero-crossing is observed~\cite{proposal_b1_jlab}. The expected result from the approved $b_1$ experiment considering 85 nA electron beam incident on $26\%$ enhanced tensor polarized target for 36 PAC days, along with different theoretical models is shown in Fig.~\ref{fig:b1_projected}~\cite{proposal_b1_jlab_jeopardyPAC51}.}

Additionally, a CAA proposal~\cite{CAA_tensor_arxiv} was recently submitted to analyze the CLAS12 Run Group-C (RGC) experiment, which took data during 2022 and 2023 in Hall-B of Jefferson Lab. This CAA proposal focuses on the analysis of existing CLAS12 data collected by impinging electrons on the longitudinally polarized deuterated ammonia target. The analysis of inclusive events to extract the tensor structure function $b_1$ will enable the early measurement of this structure function at Jefferson Lab, serving as a preparation for the dedicated $b_1$ experiment in Hall-C. The RGC was not an experiment dedicated to a tensor target, so information on the tensor polarization ($\mathcal{Q}$) of the target is calculated using the available vector (longitudinal) polarization as $\mathcal{Q} = 2- \sqrt{4-3\mathcal{P}^2}$, which is valid at thermal equilibrium in the solid lattice. However, further analysis of the RGC nuclear magnetic resonance spectra, utilizing deep neural networks, is underway to reduce the target polarization uncertainty.

\subsection{SIDIS Programs}

With the CAA proposal, we have also proposed to analyze SIDIS events with pions in the final state $d(e,e'\pi^\pm)X$~\cite{CAA_tensor_arxiv}, utilizing the CLAS12 RGC data on the vector polarized deuterated ammonia. Transverse-momentum-dependent tensor structure functions of deuteron has never been experimentally studied before, and this analysis will be an exploratory study of these tensor contributions to the SIDIS cross-section. Considering only the longitudinal polarization of tensor target, the SIDIS formalism is presented in Sec.~\ref{sec:2b_sidis}. We will extract the tensor TMD structure functions expressed in Eq.~\ref{eq:TMD_observable} with single pion SIDIS reactions $d(e,e'\pi^+)X$. We will also analyze another channel $d(e,e'\pi^-)X$ to access these tensor structure functions. \revisedJP{These two channels contain distinct information of quark distributions convoluted with fragmentation functions. Analysis of both channels helps to isolate the partonic information by combining the results appropriately. At large $x$, ratio of valence quark distribution is accessible via these $\pi^+$ and $\pi^-$ flavor tagged SIDIS processes. Furthermore, both type of SIDIS events are accessible via same set of CLAS12 data, so we will look both channels and extract results on each flavor tagged ($\pi^+$ and $\pi^-$) process.} The selected events will have one scattered electron and pion in the final state, both detected in the forward detector system of the CLAS12 spectrometer which covers the polar angle from $5^\circ$ to $35^\circ$, and almost $360^\circ$ azimuthally~\cite{CLAS12_BURKERT}. CLAS12 has a large coverage on $Q^2$, $x$, $z$ and $P_{h\perp}$ with 11.6 GeV beam and we look events in a wide kinematic range: $Q^2 > 0.95$~(GeV/c)$^2$, $0 < P_{h\perp} < 0.8$~GeV/c, $0.08 < x < 0.8$ and $0.2 < z < 0.7$. Despite the lower tensor polarization during the RGC, average about 10\%,  analysis of single pion SIDIS events allows to extract the tensor TMD structure functions of the deuteron. This measurement will provide both novel results and strong motivation for future dedicated experiments.
\begin{figure*}[htb]
    \centering
    \includegraphics[height=8cm,width=11cm]{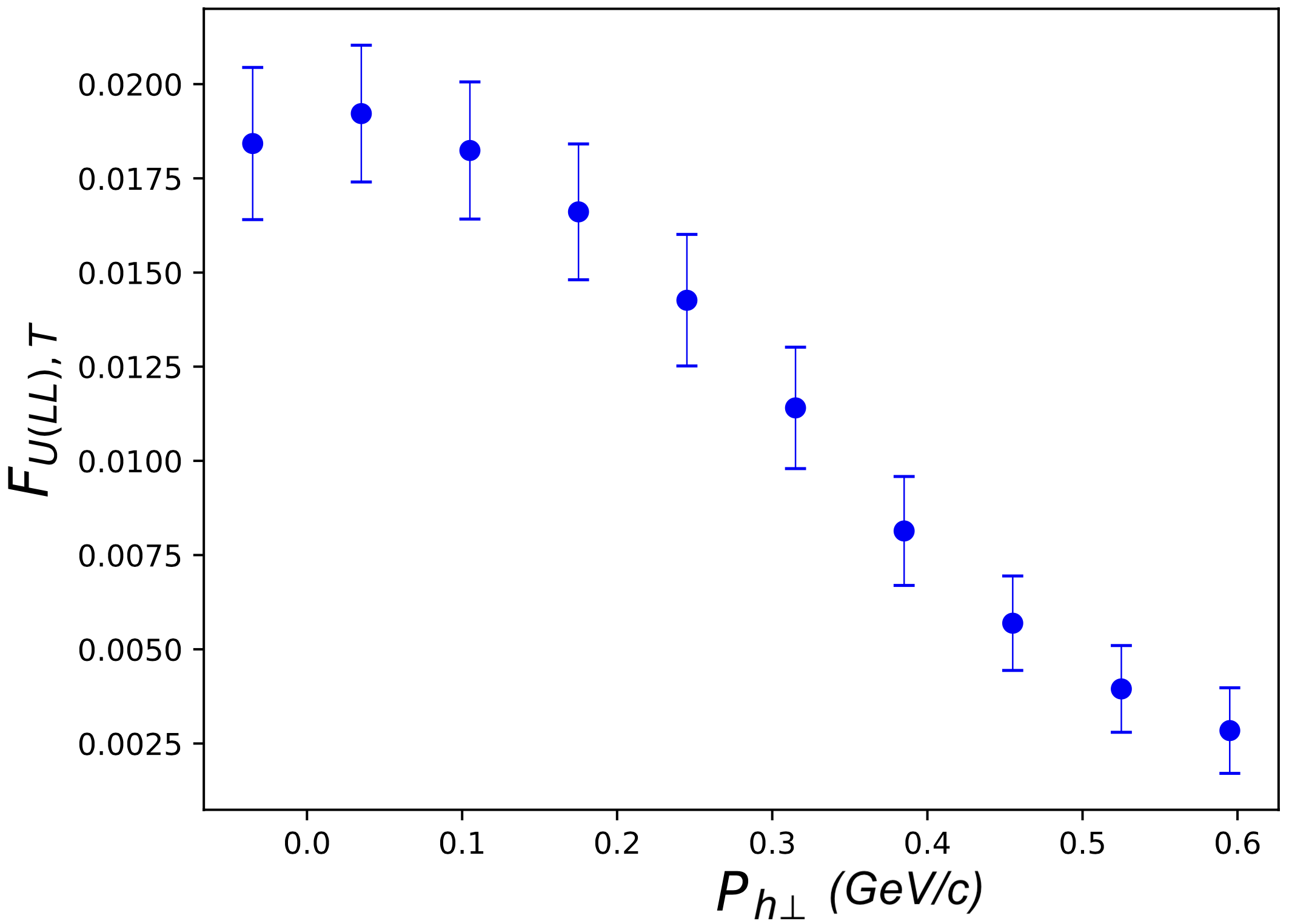}
    \caption{Projected estimation of tensor structure function $F_{U(LL),T}$ from simulation considering a constant 10\% scaling of unpolarized structure function $F_{UU,T}$ as a function of transverse momentum of detected final state \revisedJP{$\pi^+$ with the selected kinematic region of $1.0<Q^2<2.5$ (GeV/c)$^2$, $0.1<x<0.35$ and $0.3<z<0.7$.  Error bars are estimated considering the 80 nA of 11 GeV electron beam impinging on $30\%$ tensor polarized deuteron target for 5 days at Hall-C~\cite{LOI_pac52_spin1}.}}
    \label{fig:FULLT_projected}
\end{figure*}

Additionally, a letter of intent (LOI) was  submitted to the Program Advisory Committee (PAC52) of Jefferson Lab to measure tensor structure functions with a dedicated experiment in the Hall-C~\cite{LOI_pac52_spin1}. Such tensor experiments are viable with the latest available techniques to enhance the target tensor polarization~\cite{Keller:2020wan,Clement:2023eun} which can achieve an average tensor polarization of $\sim 30\%$ for the experiments under beam conditions. The LOI proposes to commission the Super-BigBite-Spectrometer (SBS) in Hall-C, which provides larger acceptance, to detect final state pions in the $d(e,e'\pi^\pm)X$, and the standard SHMS detector to measure the final electrons. This measurement is proposed \revisedJP{to} collect data on different kinematic points on $x$ and $Q^2$, and can provide higher precision result on tensor TMD distributions.  The LOI estimates for $1 < Q^2 < 2.5~$(GeV/c)$^2$, $0.1 < x < 0.35$ and $0.3 < z < 0.7$, with an electron-beam of 11~GeV. These kinematic settings could change when optimizing for the full proposals to account for all experimental and physics requirements.

The expected tensor structure function $F_{U(LL),T}$ proposed in the LOI, as shown in Fig.~\ref{fig:FULLT_projected}, was generated using \revisedJP{simulated data in a single kinematic setting of Super High Momentum Spectrometer (SHMS) and Super Bigbite Spectrometer (SBS) at Hall-C}, and considering a constant 10\% scaling of unpolarized tensor structure function $F_{UU,T}$. The 10\% scaling factor \revisedJP{for the projection of $F_{U(LL),T}$} was chosen based on the inclusive $b_1$ results from the HERMES compared to the unpolarized contribution, but better estimates will be done after having results from the exploratory measurement with the CAA. A full experimental proposal based on this LOI is in preparation for a dedicated tensor TMD experiment on deuteron with an enhanced tensor polarized target which will be submitted to the coming PAC.

Furthermore, future precision measurements and multidimensional studies of tensor structure functions can be expanded in the CLAS12 of Hall-B as well as in the SoLID of \revisedJP{Hall-A}. SoLID will provide a large angular and momentum acceptance~\cite{SoLID_whitePaper2024}, which will be suitable for the multidimensional study of the tensor structure functions with higher precision on various kinematics. Along with an enhanced tensor polarized target, excellent capability of the SoLID to handle higher data rate at higher luminosity will allow getting higher precision measurements of tensor structure functions.

\section{Summary}\label{sec:5}

Deuteron is the lightest nucleus with two spin-$\frac{1}{2}$ nucleons and it has been widely used to get the neutron structure function. Deuteron itself has an intriguing spin structure due to its tensor nature that has rarely been explored experimentally because of the experimental challenge on tensor polarized target. \revisedJP{Various theoretical model calculations of deuteron's structure function $b_1(x)$ are accessible, but cannot explain the existing HERMES data accurately. So, validation of these models require additional experimental data. Additionally, a covariant model calculation of leading-twist TMDs for spin-1 targets is formulated recently, which will be providing estimation of TMDs on deuteron soon~\cite{Cloet_private_2025}. The interesting features of tensor polarized TMDs that's mentioned in Ref.~\cite{PhysRevC_Cloet_2017} would be an avenue to validate with an experimental data in the future. All these} tensor distribution functions of the deuteron provides the information for fully understanding the structure of light nuclei at the partonic level. \revisedJP{These functions will provide a novel enlightenment about the interplay between the QCD dynamics and the nuclear structure, paving a more refined way of the QCD studies.}

After a significant progress on enhancing the tensor polarization of the deuteron target, the tensor studies on the deuteron is gradually expanding at Jefferson Lab. An experimental program to measure the tensor structure function $b_1$ with an improved precision compared to HERMES result, is already approved to run in Hall-C. In pursuance to explore the tensor TMDs experimentally, two programs, CAA and a dedicated Tensor TMD experiment, are currently underway at Jefferson Lab which can provide the very first outlook of the tensor TMDs on the deuteron. These tensor programs can be further expanded for the precision multidimensional studies in the SoLID  as well as \revisedJP{luminosity upgraded CLAS12} taking an advantage of larger acceptance and higher luminosity. \revisedJP{Additionally, 22 GeV upgrade of Jefferson Lab will cover much higher $Q^2$ region which will be a suitable avenue to further investigate the tensor structure functions and TMDs on deuteron. All of these experimental activities to measure the tensor components of the deuteron structure would attract and encourage the scientific community to involve more on model calculations and estimations of tensor TMDs.}

\backmatter

%\bmhead{Supplementary information}

\bmhead{Acknowledgements}
%Acknowledgements are not compulsory. Where included they should be brief. Grant or contribution numbers may be acknowledged.
This material is based upon work supported by the U.S. Department of Energy, Office of Science, Office of Nuclear Physics under contracts DE-AC05-06OR23177 and DE-SC0024665 .%work under Jefferson Lab contract and Santiesteban Grant 

%##################################
%commenting out with iffalse ..fi
\iffalse
\section*{Declarations}

Some journals require declarations to be submitted in a standardised format. Please check the Instructions for Authors of the journal to which you are submitting to see if you need to complete this section. If yes, your manuscript must contain the following sections under the heading `Declarations':

\begin{itemize}
\item Funding
\item Conflict of interest/Competing interests (check journal-specific guidelines for which heading to use)
\item Ethics approval and consent to participate
\item Consent for publication
\item Data availability 
\item Materials availability
\item Code availability 
\item Author contribution
\end{itemize}

\noindent
If any of the sections are not relevant to your manuscript, please include the heading and write `Not applicable' for that section.

\begin{appendices}

\section{Section title of first appendix}\label{secA1}

Text here



\end{appendices}
\fi
% end of comment out of iffalse .. fi

%\bibliographystyle{}
%\bibliographystyle{unsrt}
\bibliography{bibliography.bib}

\end{document}